\documentclass[useAMS,usenatbib,usegraphicx]{mn2e}

\title[Be star HR~7409]
{On the nature of the Be star HR~7409 (7~Vul)}
\author[S. Vennes et al.]{S. Vennes$^{1}$\thanks{E-mail: vennes@sunstel.asu.cas.cz (SV); kawka@sunstel.asu. cas.cz (AK); 
af07152@alas.matf.bg.ac.rs (SJ); af07171@alas. matf.bg.ac.rs (IP);
liliev@astro.bas.bg (LI); kubat@sunstel.asu. cas.cz (JK); slechta@sunstel.asu.cas.cz (MS); nemeth@sunstel. asu.cas.cz (PN); kraus@sunstel.asu.cas.cz (MK)},
A. Kawka$^{1}$, 
S. Joni\'{c}$^{2}$, 
I. Pirkovi\'{c}$^{2}$, 
L. Iliev$^{3}$, 
J. Kub\'at$^{1}$, 
M. \v{S}lechta$^{1}$,
\newauthor P. N\'emeth$^{1}$,
and M. Kraus$^{1}$\\ 
$^{1}$Astronomick\'y \'ustav AV \v{C}R, Fri\v{c}ova 298,CZ-251 65 Ond\v{r}ejov, Czech Republic\\
$^{2}$Department of Astronomy, Faculty of Mathematics, University of Belgrade, Studenski trg 16, 11000 Belgrade, Serbia\\
$^{3}$Institute of Astronomy, Bulgarian Academy of Sciences, 72 Tsarigradsko Shossee Blvd., 1784 Sofia, Bulgaria} 

\begin{document}

\date{Accepted . Received ; in original form }

\pagerange{\pageref{firstpage}--\pageref{lastpage}} \pubyear{2010}

\maketitle

\label{firstpage}

\begin{abstract}
HR~7409 (7~Vul) is a newly identified Be star possibly part of the Gould Belt and is the massive component of a 69-day spectroscopic
binary. The binary parameters and properties of the Be star measured using high-dispersion spectra 
obtained at Ond\v{r}ejov Observatory and at 
Rozhen Observatory imply the presence of a low mass companion ($\approx 0.5-0.8\,M_\odot$). If the pair is relatively young ($<50-80$ Myr), then the companion is a K~V star, but, following
another, older evolutionary scenario, the companion is a horizontal-branch star or
possibly a white dwarf star. In the latter scenario, a past episode of mass
transfer from an evolved star onto a less massive dwarf star would be responsible for the peculiar nature of the present-day, fast-rotating Be star.
\end{abstract}


\section{Introduction}

The B5 star HR~7409 (7~Vul, BD+19 4039, TD1~24807, HD~183537, HIC/HIP~95818) was noted for its
broad helium lines and other unspecified spectroscopic peculiarities \citep{les1968}. 
The broad helium and magnesium line profiles show that HR~7409 has a high rotation velocity
$v\sin{i}= 300\pm30$\,km\,s$^{-1}$ \citep{wol1982,abt2002}.
\citet{hil1976} found it to be photometrically variable ($\Delta m\approx 0.1$ mag) but with an unknown period,
although Hipparcos photometry only shows possible variations of 0.009 mag 
and an ``unsolved'' periodicity of 0.59 d during the survey time line \citep{koe2002}.
Using the same data \citet{mol2002} measured variations of 0.013 mag with a period
of 2.72 d.
The ultraviolet (UV) spectrum obtained with the TD-1 satellite \citep{jam1976} shows a B5-type star intermediate to main-sequence 
and supergiant \citep{cuc1980}.
The UV photometric measurements obtained with the Astronomical Netherlands 
Satellite \citep[ANS,][]{wes1982} are consistent with TD-1 flux measurements \citep{tho1978} and show the effect of
interstellar reddening \citep{sav1985,fri1992} consistent with its proximity to the Galactic plane
($l=55.2026,\,b= +1.1794$). The ANS photometry also shows evidence of variability.

HR~7409 originally figured among stars assembled as a likely open cluster around 4 and 5~Vul \citep{mey1903, mey1905} 
and later known as Collinder~399 \citep{col1931}. After the Collinder~399 membership was reduced to six objects 
that excluded HR~7409 \citep{hal1970}, the increased precision of parallax and proper motion measurements (Hipparcos, 
Tycho, Tycho2) allowed \citet{bau1998} and \citet{dia2001} to conclude that Collinder~399 is not a real cluster. 
In summary, HR~7409 is not part of a cluster, but it may belong to a local system or association referred to as
the Gould Belt \citep{les1968}.

\citet{pla1921} initially listed HR~7409 as a new spectroscopic binary but additional radial velocity
measurements \citep{pla1931} contradicted their initial assessment.
\citet{pla1921} noted the ``nebulous'' helium lines while broad hydrogen lines were also noted
by \citet{pla1931}. Both observations were evidence of a high rotation velocity.
A few additional radial velocity measurements that differ from these early measurements are available 
in the literature \citep{duf1995,feh1996} underlining difficulties in measuring radial velocities in fast rotating
B stars.
On the other hand, \citet{abt1984} did not consider HR~7409 to be in a binary.

Although HR~7409 was known
to be peculiar, a detailed spectroscopic investigation of a possible link to the Be
phenomenon is, so far, lacking.
\citet{sle1976} examined the role played by rotation on the Be phenomenon \citep[see the review by][]{por2003} and concluded that although Be stars
are apparently rotating near critical velocity, other mechanisms are possibly responsible for episodes of mass loss.
In fact, \citet{fre2005} found that the average rotation rate is 88\% of the critical velocity, while
\citet{cra2005} concluded that Be stars are sub-critical rotators.

We present in Section 2 a series of H$\alpha$ high-dispersion spectra showing that HR~7409 
is a Be star in a single-lined spectroscopic binary. The H$\alpha$ emission appears variable on a time-scale
of years. We constrain the parameters of the Be star using the ultraviolet-to-optical spectral energy
distribution (Section 3.1) and Balmer line profiles (Section 3.2). Next, we measure the binary parameters
(Section 4.1) and mass function of the companion (Section 4.2) while attempting to retrace the prior evolution of the
system. We summarize and conclude in Section 5.

\section{Observations}

Table~\ref{tbl-1} lists astrometric and photometric measurements of HR~7409. 
The ultraviolet (ANS), optical (UBV), and infrared (2MASS) photometry and the
Hipparcos parallax help calculate the extinction-corrected absolute luminosity. 
The ANS magnitudes are converted to flux measurements using
\begin{displaymath}
\log f({\rm W\,m^{-2}\,nm^{-1}}) = -0.4\,(m+26.1).
\end{displaymath}

\begin{table}
\begin{minipage}{80mm}
\centering
\caption{HR~7409 (7~Vul)}
\label{tbl-1}
\renewcommand{\footnoterule}{\vspace*{-15pt}}
\begin{tabular}{ccc}
\hline
Parameter & Value & Ref.\footnote{{\it References:} (1) \citet{per1997}; (2) \citet{van2007}; (3) \citet{wes1982};
(4) \citet{cra1971}; (5) \citet{mer2006}; (6) \citet{skr2006}; (7) \citet{sav1985}; (8) \citet{fri1992};
(9) \citet{nec1980}; (10) \citet{gua1992}.} \\
\hline
\multicolumn{3}{c}{Hipparcos} \\
\hline
$\pi$ (mas)  &  4.29$\pm$0.76, 2.81$\pm$0.48 & 1,2 \\
$\mu_\alpha\cos{\delta}$ (\,mas\,yr$^{-1}$) & 2.95$\pm$0.46, 1.38$\pm$0.43 & 1,2 \\
$\mu_\delta$ (\,mas\,yr$^{-1}$) & $-$16.89$\pm$0.54, $-$15.84$\pm$0.55 & 1,2\\
       \hline
  \multicolumn{3}{c}{ANS photometry} \\
       \hline
$m$[155$\pm$15\,nm] (mag) & 4.261$\pm$0.014 & 3 \\
$m$[180$\pm$15\,nm] (mag) & 4.472$\pm$0.018 & 3 \\
$m$[220$\pm$20\,nm] (mag) & 4.928$\pm$0.014 & 3 \\
$m$[250$\pm$15\,nm] (mag) & 5.116$\pm$0.011 & 3 \\
$m$[330$\pm$10\,nm] (mag) & 5.537$\pm$0.010 & 3 \\
  \hline
  \multicolumn{3}{c}{Johnson photometry} \\
  \hline
$U-B$ (mag) &  $-$0.53$\pm$0.02, $-$0.525$\pm$0.063 & 4,5 \\
$B-V$ (mag) &  $-$0.10$\pm$0.01, $-$0.107$\pm$0.012  & 4,5 \\
$V$ (mag)   &  6.31$\pm$0.02, 6.338$\pm$0.028  & 4,5 \\
  \hline
  \multicolumn{3}{c}{2MASS photometry} \\
  \hline
$J$ (mag)  &  6.439$\pm$0.026 & 6  \\
$H$ (mag)  &  6.505$\pm$0.017 & 6  \\
$K$ (mag)  &  6.554$\pm$0.024 & 6  \\
  \hline
  \multicolumn{3}{c}{Colour excess} \\
  \hline
$E_{B-V}$ (mag) & 0.06, 0.07, 0.09, 0.08 & 7,8,9,10 \\
\hline
\end{tabular}
\end{minipage}
\end{table}

We observed HR~7409 using the coud\'e spectrograph attached to the 2m
telescope at Ond\v{r}ejov Observatory \citep{sle2002}. We obtained the spectroscopic series using the 830.77
lines per mm grating with a SITe $2030\times 800$ CCD, with the slit width set at $0.7\arcsec$, resulting in 
a spectral resolution of $R = 13\,000$ and a spectral range from 6254 to 6763 \AA.
We verified the stability of the wavelength scale by measuring the wavelength centroids
of O{\sc i} sky lines. The velocity scale remained stable within 1\,km~s$^{-1}$.
Table~\ref{tbl-2} presents our observation log.

\begin{table*}
\centering
\caption{Observation log}
\label{tbl-2}
\begin{tabular}{rrrrrrrrrrrr}
\hline
UT Date & UT Start & $t_{\rm exp}$ & UT Date & UT Start & $t_{\rm exp}$ & UT Date & UT Start & $t_{\rm exp}$ & UT Date & UT Start & $t_{\rm exp}$\\
           &          &  (s) &             &          &  (s) &             &          &   (s)  &              &          & (s) \\
\hline
2007-05-23 & 22:27:44 & 1800 &   2007-07-25 & 20:13:33 & 2722 &   2009-04-13 & 01:58:08 & 2200 &   2010-04-28 & 02:05:12 &  900 \\  
2007-07-04 & 23:11:43 & 3083 &   2007-07-26 & 20:28:09 & 1800 &   2009-07-30 & 22:22:44 & 1200 &   2010-04-28 & 02:20:50 &  900 \\  
2007-07-06 & 20:31:48 & 1200 &   2007-07-27 & 23:47:42 & 3209 &   2009-07-30 & 22:43:52 & 1200 &   2010-06-26 & 00:02:11 & 1200 \\  
2007-07-07 & 01:08:58 & 1203 &   2007-08-05 & 00:12:08 & 2799 &   2009-07-31 & 20:40:23 & 1180 &   2010-06-29 & 20:49:26 &  600 \\  
2007-07-08 & 01:08:57 & 1200 &   2007-08-06 & 00:19:45 & 3000 &   2009-08-02 & 02:11:07 &  900 &   2010-06-29 & 21:07:20 &  600 \\  
2007-07-13 & 23:17:35 & 1200 &   2007-08-06 & 20:35:40 & 1200 &   2009-08-30 & 19:49:42 & 1800 &   2010-08-20 & 19:29:20 &  600 \\  
2007-07-14 & 02:10:38 &  800 &   2007-08-12 & 21:33:52 & 3200 &   2009-08-30 & 23:09:56 & 1800 &   2010-08-20 & 19:40:03 &  600 \\  
2007-07-14 & 20:38:12 & 2270 &   2007-10-13 & 17:28:29 & 3100 &   2009-09-21 & 20:18:39 & 1800 &   2010-08-21 & 19:44:16 &  600 \\  
2007-07-15 & 00:51:18 & 1426 &   2007-10-13 & 20:41:59 & 3200 &   2009-09-23 & 18:05:36 & 1200 &   2010-08-21 & 19:55:02 &  600 \\  
2007-07-15 & 01:18:16 & 1528 &   2007-10-15 & 17:32:02 &  301 &   2009-09-23 & 22:03:55 & 1200 &   2010-08-22 & 19:55:24 &  600 \\  
2007-07-15 & 01:46:26 & 1207 &   2007-10-15 & 17:38:08 & 2506 &   2009-09-23 & 22:26:49 & 1200 &   2010-08-22 & 20:06:25 &  600 \\  
2007-07-15 & 21:19:03 & 3000 &   2007-10-15 & 20:33:06 & 2900 &   2009-10-09 & 18:13:24 & 1800 &   2010-09-21 & 19:26:24 &  600 \\  
2007-07-15 & 22:12:23 & 3000 &   2007-10-16 & 18:27:36 & 2200 &   2009-10-09 & 20:52:25 & 1800 &   2010-09-22 & 19:10:45 &  600 \\  
2007-07-16 & 00:41:30 & 3000 &   2008-07-02 & 00:59:40 &  900 &   2010-04-24 & 02:16:39 & 1200 &   2010-09-23 & 18:48:06 &  900 \\  
2007-07-16 & 01:34:34 & 2700 &   2008-08-28 & 00:26:07 &  600 &   2010-04-24 & 02:37:23 & 1200 &   2010-09-24 & 18:46:04 &  900 \\  
2007-07-22 & 21:14:23 & 3684 &   2008-10-13 & 18:22:53 &  600 &   2010-04-25 & 02:45:10 &  900 &   2010-09-30 & 18:34:54 &  900 \\  
2007-07-22 & 23:29:54 & 4309 &   2008-10-13 & 18:36:29 & 2200 &   2010-04-26 & 02:37:15 &  900 &   2010-10-13 & 18:58:08 &  600 \\
\hline
\end{tabular} 
\end{table*}

We also obtained a series of spectra on UT 2010 Nov 15 using the coud\'e spectrograph
attached to the 2m telescope at Rozhen National Astronomical Observatory (Bulgaria).
We used the 632 lines per mm grating with a SITe $1024\times1024$ CCD. We set the slit width
at $0.83\arcsec$ resulting in a
spectral resolution of $R = 31\,000$, 20\,000, 17\,000, and 16\,000 at four tilt angles centred on H$\alpha$,
H$\beta$, H$\gamma$, and H$\delta$, respectively.

All spectra were wavelength calibrated with a ThAr comparison arc spectra obtained shortly after
each exposure. The telluric features in the H$\alpha$ spectra were removed using a fast-rotating B star template (HR~7880).
The data were reduced using standard IRAF procedures.

The first spectrum obtained on 2007 May 23 revealed a fast rotating star with a
H$\alpha$ emission core typical of Be stars. Spectra obtained in the following months and years
show the emission component decreasing in strength. The Ond\v{r}ejov spectra also show the rotationally
broadened lines He\,{\sc i}$\lambda$6678, Si\,{\sc ii}$\lambda\lambda$6347.109,6371.371,
and Ne\,{\sc i}$\lambda$6402.246.

\section{Properties of the B\lowercase{e} star}

We analysed the spectral energy distribution (SED) and line profiles using a grid ($T_{\rm eff} \ge 15\,000$ K) of line-blanketed
spectra in non-local thermodynamic equilibrium \citep[non-LTE ``BSTAR'' grid,][]{lan2007}.
We supplemented this grid with models ($T_{\rm eff} \le 15\,000$ K) from a grid of spectra in local-thermodynamic equilibrium \citep[LTE,][]{cas2003}.
We selected spectra with solar abundances from both model grids.
However, the effect of gravitational darkening and geometric distortion on parameter measurements of fast-rotating B stars
such as HR~7409 are sizeable. \citet{fre2005} and \citet{lov2006} investigated the effect of near critical equatorial rotation
velocity on mean effective temperature and surface gravity measurements. The magnitude of the effect
depends on the projection angle as well as the fraction of the critical velocity attained ($v/v_c$), and may amount
to an underestimation of the temperature by 10\% and surface gravity by 0.2 dex.

We estimated the mean stellar parameters by (1) fitting the infrared to ultraviolet SED, and (2) fitting the Balmer line
series from H$\alpha$ to H$\delta$.

\subsection{Spectral energy distribution}

The SED is affected by extinction in the interstellar medium.
The colour-excess measurements range from
$E_{B-V}=0.06$ to $0.09$ mag (Table~\ref{tbl-1}), i.e., $A_V=0.19$ to 0.28 mag assuming $R_V=3.1$.
The two lowest measurements ($E_{B-V}=0.06-0.07$) were obtained by modelling the 2200\AA\ bump in the five-channel ANS ultraviolet 
photometry. The two highest measurements ($E_{B-V}=0.08-0.09$) were obtained assuming $(B-V)_0=-0.18$ mag for a B5~V type star.
HR~7409 lies toward the Vulpecula rift, an extinction wall associated with the Vulpecula molecular cloud and
rising at a distance of 0.3 kpc \citep{fre1999}, and in a region of colour-excess $E_{B-V}\approx 0.1-0.5$. 

\begin{figure}
\includegraphics[width=1.00\columnwidth]{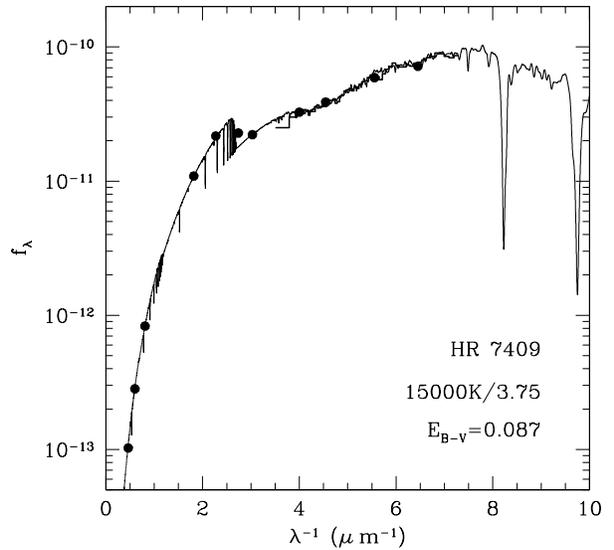}
\caption{Spectral energy distribution built using from left to right (full circles)
2MASS K, H, and J, optical V, B, and U, and ANS 330, 250, 220, 180,
and 155 nm photometric data points. The near to far UV spectrum is also
covered with TD1 spectrophotometry (thick line). The observed distribution is compared
to a model spectrum at $T_{\rm eff}=15\,000$\,K, $\log{g}=3.75$, and solar
metallicity (thin line) attenuated by interstellar extinction in the line-of-sight
($E_{B-V}=0.087$).
}\label{fig1}
\end{figure}

We fitted the SED with model spectra assuming $\log{g}=3.75$ and variable temperature and extinction. The SED comprises eleven
data points (Table~\ref{tbl-1}) with equal weights assigned to them. 
We employed the parametrized extinction curves ($R_V=3.1$) from \citet{car1989}.
A minimum effective temperature $T_{\rm eff}\ga12\,900$\,K is obtained by setting $E_{B-V}=0$. 
Allowing the temperature to increase to 14\,000, 15\,000, and 16\,000\,K, we found
that the best-fitting $E_{B-V}$ increased to 0.053, 0.087, and 0.117, respectively.
Varying the temperature and the colour-excess simultaneously the best-fitting parameters are:
\begin{displaymath}
T_{\rm eff}=14\,400\pm800\,{\rm K},\ E_{B-V}=0.069\pm0.030
\end{displaymath}
The LTE and non-LTE models at 15\,000\,K delivered consistent solutions for the colour-excess,
$E_{B-V}=0.089$ and 0.087, respectively.
Figure~\ref{fig1} shows the infrared to ultraviolet SED of HR~7409 using the non-LTE model
fit at the lower edge of the non-LTE grid (15\,000\,K).

The extinction-corrected $V$ magnitude is $V_0=6.10\pm0.10$. Therefore, 
adopting the revised Hipparcos parallax of \citet{van2007},
the distance modulus is
\begin{displaymath}
m-M = 5\log{d}-5 = 7.76_{-0.34}^{+0.41},
\end{displaymath}
and the de-reddened absolute V magnitude is $M_V=-1.66_{-0.51}^{+0.44}$, somewhat brighter than for a normal B4~V star ($M_V=-1.2$). 
\citet{lam1997} found that absolute visual magnitudes of O and B stars based on Hipparcos parallaxes correlate with
rotation velocity and may be up to 1.5 mag brighter than inferred from the apparent spectral types.

\subsection{Line profile analysis: Balmer lines}

\begin{figure}
\includegraphics[width=1.00\columnwidth]{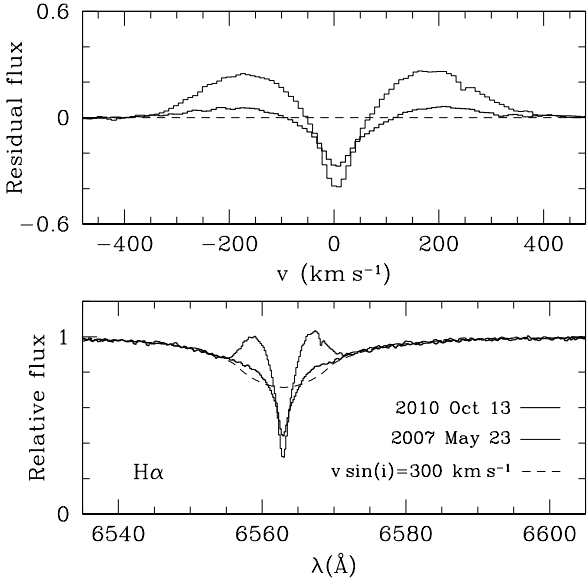}
\caption{(Bottom) optical spectra obtained at two epoches separated by $\sim$1239 days (thick and thin lines)
compared to a model at $T_{\rm eff}=15\,000$\,K and $v\,\sin{i}=300$\,km\,s$^{-1}$ (dashed line). The H$\alpha$
line profile shows broad photospheric line wings, and variable disc emission/absorption. (Top) The flux residual show
the emission and absorption components typical of a rotating disc with negligible radial motion.
}\label{fig2}
\end{figure}

\begin{figure}
\includegraphics[width=1.00\columnwidth]{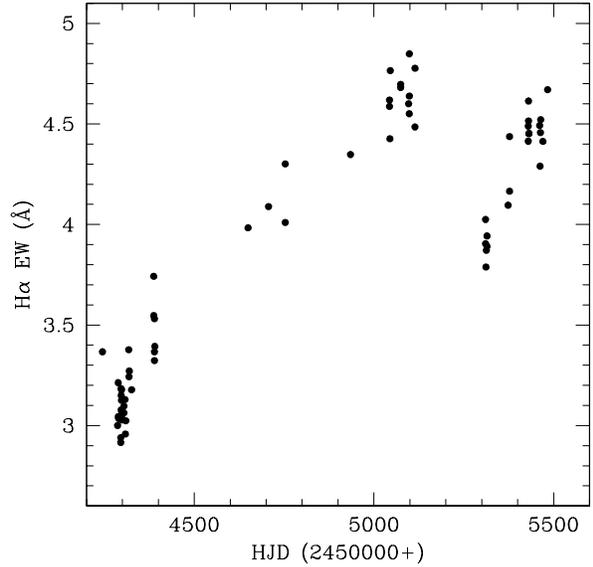}
\caption{Equivalent width (EW) of the H$\alpha$ line profile measured within a window of $\pm25$
\AA. The broad line wings are blended with an emission component shown in Figure~\ref{fig2};
the emission decreases with time resulting in increasing EW measurements.
}\label{fig3}
\end{figure}

Figure~\ref{fig2} shows H$\alpha$ spectra obtained 3.4 years apart and illustrating extreme ranges in observed line profiles.
Figure~\ref{fig3} shows the evolution
of the line equivalent width (EW) during that period.
The EW was measured using a $\pm25$ \AA\ window; the weakening of the emission wings 
over time results into deeper absorption and larger EW.
HR~7409 was caught during a declining phase. Because
disc variability may occur on time scales of years \citep{cla2003} or decades \citep{hub2007}, continuing monitoring should
assist us in capturing the next activity episode. The absence of central emission, but, instead, the
broad shoulder and narrow line core suggest the onset of a ``shell'' phase \citep[see][]{ste1999}.

\begin{figure}
\includegraphics[width=1.00\columnwidth]{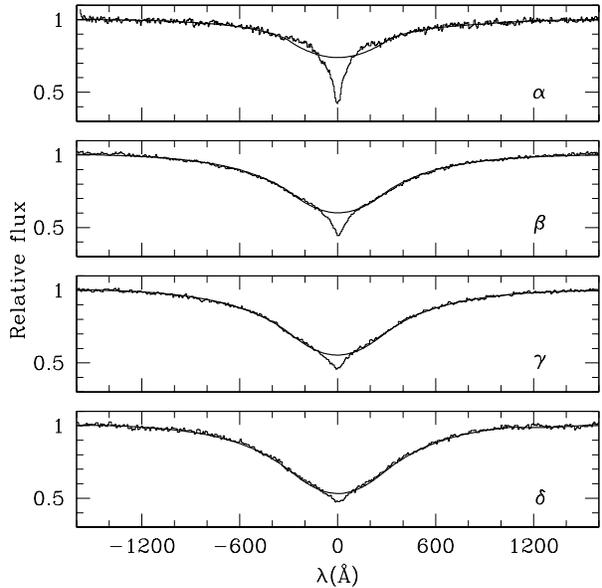}
\caption{Spectra of the Balmer line series (H$\alpha$ to H$\delta$) obtained at Rozhen and best model fit 
at $T_{\rm eff}=15\,600$~K, $\log{g}=3.75$, and $v\,\sin{i}=300$\,km\,s$^{-1}$.
}\label{fig4}
\end{figure}

Next, we fitted the Balmer line spectra observed at Rozhen Observatory using the ``BSTAR'' model grid with He$/$H$=0.1$.
The model line profiles were convolved with a rotational broadening function with $v\,\sin{i}=300$\,km\,s$^{-1}$
and limb darkening coefficient $\mu=0.4$ \citep[see][]{cra2005}. As discussed earlier, we neglected the effect of gravity darkening.
Both model and spectra are normalized at $\pm1500$\,km\,s$^{-1}$ and we excluded the line cores ($\pm400$\,km\,s$^{-1}$) from the fit.
Figure~\ref{fig4} shows the best model-fit to the Balmer line series (H$\alpha$ to H$\delta$):
\begin{displaymath}
T_{\rm eff}=15\,600\pm200\,{\rm K},\ \log{g}=3.75\pm0.02
\end{displaymath}

Parameter estimations based on the Balmer line wings and on the SED are both possibly affected by 
the fast rotation of the star. The small statistical errors in the measurements based on the
line profiles are almost certainly underestimating the true errors and do not encompass
systematic effects. Although the apparent parameters are estimated conservatively as
\begin{displaymath}
T_{\rm eff}=15\,500\pm1000\,{\rm K},\ \log{g}=3.75\pm0.13, 
\end{displaymath}
and are marginally consistent with the published spectral type B5~V, the true spectral-type is possibly earlier than this
by at least one subtype. We propose the classification B4-5\,III-IVe.
\citet{mol2002} measured $T_{\rm eff}=14\,930$ and $\log{g}=3.52$ using ubvy data, in agreement with our results.
Following the isochrones of \citet{sch1992} for $Z=0.02$ models in the ($T_{\rm eff},\log{g}$) plane as well as the
($T_{\rm eff},M_V$) plane, HR~7409 is a $5.5\pm0.5\,M_\odot$ star with an age of $\approx$50-80 Myr old nearing the
end of its main-sequence life.
The estimated age ignores, at this stage, the possibility of past binary interaction.

\section{Binary parameters and nature of the companion}

\begin{figure}
\includegraphics[width=1.00\columnwidth]{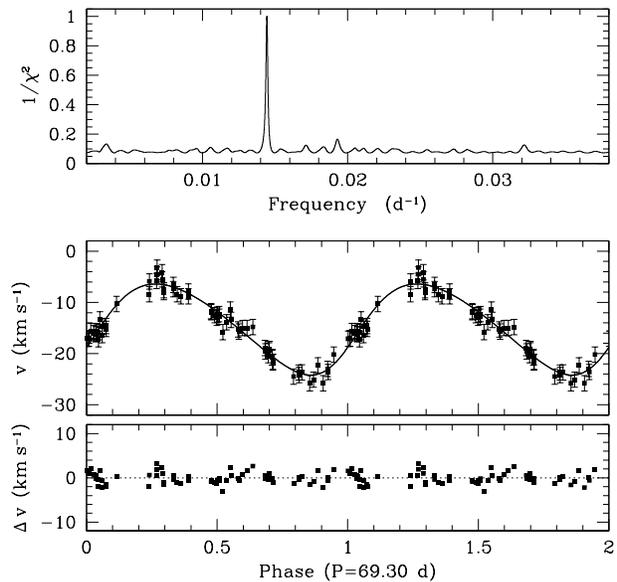}
\caption{(Top) periodogram of the H$\alpha$ line velocity measurements. (Middle) the velocity variations 
are well matched by an eccentric orbit of 69 days with (bottom) velocity residuals of only 1.3\,km\,s$^{-1}$.
}\label{fig5}
\end{figure}

We measured radial velocities using narrow H$\alpha$ line core showing, as suspected in earlier investigations, that HR~7409 resides in
a binary (Table~\ref{tbl-3}). The velocities were obtained by fitting Voigt profiles to the central 10 pixels
and we applied the heliocentric velocity correction.
The H$\beta$, H$\gamma$, and H$\delta$ velocities measured at Rozhen at a single epoch differ from the H$\alpha$ velocity by up to
$\approx20$\,km\,s$^{-1}$. The zero point of the velocity scale appears uncertain although all measurements obtained with
H$\alpha$ are internally consistent. 
Figure~\ref{fig5} shows the periodogram and H$\alpha$ radial velocity measurements phased on the orbital period.
The velocity residual is 1.3\,km\,s$^{-1}$ and is commensurate with the expected velocity accuracy.
In the following we identify the Be star with the subscript ``A'' and the unseen companion with ``B''.

\subsection{Binary parameters}

We fitted the H$\alpha$ radial velocity measurements and simultaneously constrained the 
systemic velocity $\gamma_A\equiv \gamma({\rm H}\alpha)$,
the velocity semi-amplitude $K_A\equiv K({\rm H}\alpha)$, the eccentricity $e$, and
the angle of passage of star A at $T_0$ (passage at periastron).
Table~\ref{tbl-4} lists the best-fitting orbital parameters.

Adopting $M=5.5\,M_\odot$ and $R=5.2\,R_\odot$, i.e., $\log{g}=3.75$, the critical velocity is 
$v_c=367$\,km\,s$^{-1}$ \citep[assuming $R=R_{\rm polar}$, see][]{cra2005}. Therefore, the measured rotation
velocity limits the inclination to $\sin{i}\ga 0.83$, or $i\ga 56^\circ$.

\begin{table*}
\centering
\caption{Radial velocity (H$\alpha$ core)}
\label{tbl-3}
\begin{tabular}{cccccccc}
\hline
HJD & $\textsl{v}_{\rm H\alpha}$& HJD   &  $\textsl{v}_{\rm H\alpha}$   & HJD   &  $\textsl{v}_{\rm H\alpha}$ & HJD   &  $\textsl{v}_{\rm H\alpha}$  \\ 
(2450000+)      & (km\,s$^{-1}$)&  (2450000+)      & (km\,s$^{-1}$) &  (2450000+)      & (km\,s$^{-1}$) &  (2450000+)      & (km\,s$^{-1}$) \\ 
\hline
4244.44879  & $-20.2$ &   4307.36287  & $-25.8$ &   4934.59438  & $-25.8$ &   5314.59285  & $-8.8$   \\ 
4286.48853  & $-13.4$ &   4308.36767  & $-25.1$ &   5043.44372  & $-11.7$ &   5314.60371  & $-7.9$   \\ 
4288.36662  & $-15.3$ &   4309.51439  & $-22.3$ &   5043.45840  & $-11.9$ &   5373.51247  & $-8.4$   \\ 
4288.55912  & $-15.7$ &   4317.52888  & $-17.1$ &   5044.37252  & $-12.5$ &   5377.37525  & $-8.0$   \\ 
4289.55911  & $-15.0$ &   4318.53531  & $-15.6$ &   5045.60056  & $-12.3$ &   5377.38768  & $-7.6$   \\ 
4295.48184  & $-18.9$ &   4319.36927  & $-16.0$ &   5074.33998  & $-23.0$ &   5429.31932  & $-17.2$  \\  
4295.59970  & $-19.6$ &   4325.42111  & $-10.3$ &   5074.47902  & $-23.6$ &   5429.32676  & $-15.8$  \\  
4296.37736  & $-20.3$ &   4387.24668  & $-17.1$ &   5096.35883  & $-5.9$  &   5430.32966  & $-16.4$  \\  
4296.54824  & $-19.1$ &   4387.38162  & $-17.1$ &   5098.26284  & $-5.8$  &   5430.33713  & $-14.7$  \\  
4296.56756  & $-19.1$ &   4389.23280  & $-16.2$ &   5098.42833  & $-4.5$  &   5431.33735  & $-14.6$  \\  
4296.58526  & $-19.4$ &   4389.24980  & $-15.6$ &   5098.44423  & $-3.2$  &   5431.34500  & $-15.2$  \\  
4297.40996  & $-20.8$ &   4389.37358  & $-15.7$ &   5114.27061  & $-13.0$ &   5461.31562  & $-12.8$  \\  
4297.44700  & $-20.7$ &   4390.28230  & $-13.3$ &   5114.38103  & $-12.4$  &  5462.30469  & $-15.8$  \\  
4297.55055  & $-21.7$ &   4649.55082  & $-24.5$ &   5310.60225  & $-7.4$   &  5463.29063  & $-13.9$  \\  
4297.58567  & $-21.6$ &   4706.52511  & $-15.0$ &   5310.61665  & $-6.6$   &  5464.28915  & $-11.4$  \\  
4304.41070  & $-23.7$ &   4753.26994  & $-4.1$  &   5311.62039  & $-8.4$   &  5470.28099  & $-14.8$  \\  
4304.50842  & $-24.2$ &   4753.28864  & $-5.5$  &   5312.61496  & $-8.8$   &  5483.29445  & $-23.7$  \\ 
\hline
\end{tabular} 
\end{table*}

\subsection{Nature of the companion and evolutionary scenarios}

We may now constrain the nature of the binary companion by calculating the mass
function:
\begin{displaymath}
f(M_B)=\frac{P\,K_A^3}{2\pi G}(1-e^2)^{3/2} = (4.9\pm0.7)\times10^{-3}\,M_\odot,
\end{displaymath}
and solving iteratively for the secondary mass $M_B$:
\begin{displaymath}
M_B = f(M_B) \frac{(1+q)^2}{\sin^3{i}},
\end{displaymath}
where $q=M_A/M_B$. The orbit and the Be star rotation are almost certainly co-planar. Adopting $\sin{i}=0.83$ for the inclination of the Be star rotation plane, i.e., assuming sub-critical rotation velocity, and adopting $M_A=5.5\pm0.5\,M_\odot$,
we find
$M_B = 0.62-0.77\,M_\odot$ and $q=7.8-8.1$. A lower limit on the secondary mass is set by assuming
$\sin{i}=1$  resulting in
$M_B=0.50-0.63\,M_\odot$, or $q=9.5-10$. 
In summary, the secondary star has a mass within the range $M_B=0.50-0.77\,M_\odot$ for a binary mass ratio $q=7.8-10$.
The semi-major axis is $a=130\pm5\,R_\odot$ or $0.60\pm0.02$ au.

The mass of the companion is typical of M1 to K1 main-sequence stars, but also of the bulk of white dwarf stars \citep{shi1979}.
If the companion is a main-sequence star, then the system is relatively young \citep[young-scenario: 50-80 Myr;][]{sch1992} and with a total
systemic mass of $5.5-6.8\,M_\odot$. If the companion is a white dwarf then the current orbital separation necessarily implies past
interaction and we must investigate plausible evolutionary scenarios \citep[old-scenario, see][]{wil2004}. The outcome of the old-scenario is
not necessarily a white dwarf plus main-sequence binary, but the evolved component of the system may also be caught at shorter-lived, intermediary
stages.

The mass-accreting component of a close binary may acquire sufficient angular momentum to reach critical rotation velocity \citep{pac1981}.
Although it was originally proposed that Be stars are subjected to ongoing accretion \citep{kri1975}, the properties of 
Be stars rather suggest that many, but not all, are exhibiting the effect of past rather than current accretion events, more specifically a case-B mass
transfer events while the evolving star climbed the giant branch \citep{wat1989,pol1991,van1997}.
Indeed, \citet{wat1991} find evidence that the Be binary HR~2142 holds a white dwarf or He-star secondary, and
\citet{gie1998} show that the companion to the Be star $\phi$ Per \citep{poe1981} is an extreme-horizontal branch (EHB) star.
HR~7409 shares a number of characteristics with the Be star 4~Her \citep{kou1997}. Both stars
are fast rotating and reside in long period binaries. Moreover, their mass functions
imply the presence of a low-mass companion. However, our interpretation of the 
phenomenon involves past interaction and mass transfer rather than on-going
accretion as proposed by \citet{kou1997}.

Population syntheses \citep[see, e.g.,][]{rag2001,wil2004} aim at predicting the binary period and final-mass distributions.
\citet{wil2004} described a likely scenario (labelled number ``2'') for the formation of the present-day binary HR~7409: a $5.5+3.5\,M_\odot$ pair in
a close 5-day binary is expected to experience its first Roche lobe overflow (RLOF) event after the primary crosses the Hertzsprung gap and climbs
the giant branch ($t\approx 80$ Myr). The process quickly results in a reversal of the mass ratio (from $q=1.6$ to $q=0.12$). 
With the cessation of mass transfer, the binary enters an extended quiet period while the evolved star, labelled a
``naked'' helium star, sits on the horizontal branch (HB) for $\approx 20$ Myr. After exhaustion of the helium fuel, the binary
enters a second RLOF episode and becomes a detached white dwarf plus B-star binary with a final period of 93 days.
This scenario is not only applicable to the formation of Be stars but to any fast-rotating B stars. \citet{gie2008}
found that the fast-rotating B star Regulus is a 40-day binary with a likely white dwarf companion.
\citet{rap2009} reconstructed the past history of this system and concluded that the Regulus system will likely
evolve into an AM CVn system.

HR~7409 appears to be a less massive version of scenario number 2 described by \citet{wil2004}.
Assuming conservative mass-transfer throughout the evolution we may assume initial masses of
$M_{1,0}=3.4$ \citep[$t_{\rm MS}=250$ Myr,][]{sch1992} and $M_{2,0}=2.7\,M_\odot$ for the original primary and secondary and an
orbital period of several days. In this scenario, $\approx2.8\,M_\odot$ are transfered from the
original primary to the secondary leaving a fast-rotating $\approx5.5\,M_\odot$ star with an evolved $\approx0.6\,M_\odot$
He-star or white dwarf companion. Considering that the present day Be star exhibits evidence of a wind, an
accreting white dwarf would emit copious X-rays as in the case of $\gamma$ Cas \citep{har2000}
or HD~161103 and SAO~49725 \citep{lop2006}. However,
HR~7409 is not an X-ray source \citep{ber1996}, leaving us with the possibility that the unseen companion
is a He-star with a shallower potential well.

\begin{table}
\begin{minipage}{80mm}
\centering
\caption{Orbital parameters}
\label{tbl-4}
\renewcommand{\footnoterule}{\vspace*{-15pt}}
\begin{tabular}{cc}
\hline
Parameter & Value  \\
\hline
$P$  & $69.30\pm0.07$\,d   \\
$T_0$ & HJD~$2454248.1\pm2.7$ \\
$\gamma$(H$\alpha$) &  $-14.8\pm0.2$\,km\,s$^{-1}$ \\
$K$(H$\alpha$)      & $8.9\pm0.4$ \,km\,s$^{-1}$\\
$\omega$  & $247\pm16^\circ$ \\
$e$       & $0.161\pm0.035$ \\
\hline
\end{tabular}
\end{minipage}
\end{table}

Having been striped of its hydrogen envelope, the naked He-star may join other HB stars with $M_V\approx 0.5$ and
contribute up to 15\% of the composite optical flux, or
resemble a hot subdwarf on the EHB with
absolute V magnitude $\approx 4$ \cite[see][]{heb2009} and would contribute less than 1\% of the composite flux.
Further examination of the high-dispersion spectra may offer clues to the nature of the companion.

\subsection{Line profile analysis: He\,{\sc i}$\lambda$6678.15}

\begin{figure}
\includegraphics[width=1.00\columnwidth]{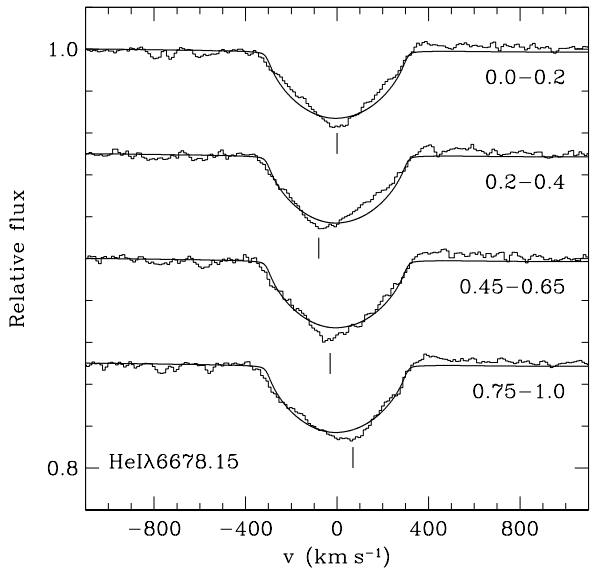}
\caption{Co-added He\,{\sc i}$\lambda$6678 spectra phased with the orbital period. The velocity scale is set in the Be star rest frame. The line asymmetry varies with
the orbital phase. The spectra labelled $0.2-0.4$, $0.45-0.65$, and 0.75-1.0 are shifted down by 0.05, 0.10, and 0.15, respectively.
}\label{fig6}
\end{figure}

Figure~\ref{fig6} shows co-added He\,{\sc i}$\lambda$6678 spectra phased with the orbital period.
In order to increase the signal-to-noise ratio and search for weak spectral lines tracing the orbit of the
companion we regrouped the spectra in four bins and adjusted the velocity scale to the Be star rest-frame.
The He\,{\sc i}$\lambda$6678 line shows a broad red-shifted emission and its overall shape varies over time.
A weak absorption feature ($\approx 30$\,m\AA) marked with vertical lines is moving in opposite directions to the Be star 
($v(0.0-0.2,0.2-0.4,0.45-0.65,0.75-1.0)\approx-15.,-85.,-45.,45.$\,km\,s$^{-1}$) and may belong to a
subluminous B-type companion. The full amplitude of the motion, after correcting for orbital smearing in the co-added spectra, is $K\approx 140$\,km\,s$^{-1}$ and would imply
a mass ratio $\approx 8$ for the companion in agreement with the mass function (Section 4.2).

A confirmation of the detection of a secondary star may be secured with optical or ultraviolet echelle
spectroscopy at high-dispersion and very high-signal-to-noise ratio.

\section{Summary and conclusions}

Table~\ref{tbl-5} lists some stellar properties.
We found that the Be star HR~7409 forms a 69.2 d binary with a low-mass companion.
Although, we cannot exclude a low-mass main-sequence star, the companion is most likely
a hot sub-luminous star. An evolutionary scenario involving an episode of conservative
mass-transfer also offers a natural explanation for the fast rotation of the Be star.
Using the distance, systemic velocity and proper-motion measurements ($d$, $v_r$, $\mu_\alpha$, and $\mu_\delta$)
we confirm the peculiar kinematics of the Be star which imply that HR~7409 belongs to a group of
runaway B stars \citep{hoo2001}. Instead, we propose that its prior evolution and ``rejuvenation'' suggests
more evolved kinematical properties such as those of white dwarf stars that show a lag in the 
Galactic $V$-component of $-40$\,km\,s$^{-1}$ \citep{sio1988}.

A relatively old age for the system would imply that it is not part of the Gould Belt. The main-sequence lifetime
of the evolved star is estimated at 250 Myr for an initial mass of 3.4\,$M_\odot$ (see Section 4.2), but increasing
the progenitor mass above 5\,$M_\odot$ ($t_{\rm MS}\la90$ Myr) would bring the system age closer to the estimated age of the Gould Belt
\citep[30-60 Myr, see][and references therein]{bek2009}. Detailed evolutionary models for this particular system should help elucidate the
age problem.

\begin{table}
\begin{minipage}{80mm}
\centering
\caption{Kinematics, distance, masses, and spectral types}
\label{tbl-5}
\renewcommand{\footnoterule}{\vspace*{-15pt}}
\begin{tabular}{cc}
\hline
Parameter & Value  \\
\hline
$(U,V,W)$  & $(6^{+5}_{-3},-42^{+2}_{-4},-9^{+3}_{-4})$\,km\,s$^{-1}$ \\
$d$  & $356^{+75}_{-51}$\, pc \\
$M_A,M_B$ & $5.5\pm0.5,\ 0.50-0.77\ M_\odot$ \\
Types & B4-5\,III-IVe plus (EHB, HB, K~V, or WD) \\
\hline
\end{tabular}
\end{minipage}
\end{table}

\section*{Acknowledgments}
S.V. and A.K. are supported by GA AV grant numbers IAA300030908 and IAA301630901, respectively, and by GA \v{C}R grant number P209/10/0967.
A.K. also acknowledges support from the Centre for Theoretical Astrophysics (LC06014). 
The visit of S.J. and I.P. at Ond\v{r}ejov Observatory was supported by the Department of Astronomy, Faculty of Mathematics,
University of Belgrade. We thank the referee, J. Zorec, for a prompt and informative review.
We also thank P. \v{S}koda, P. Koubsk\'y, M. Netolick\'y, J. Polster, B. Ku\v{c}erov\'a, D. Kor\v{c}\'akov\'a, and P. Hadrava 
for obtaining some of the spectra used in the present study.

This research has made use of the SIMBAD database, operated at CDS, Strasbourg, France.
This publication makes use of data products from the Two Micron All Sky Survey, which is a joint project of the 
University of Massachusetts and the Infrared Processing and Analysis Center/California Institute of Technology, 
funded by the National Aeronautics and Space Administration and the National Science Foundation.

\label{lastpage}

\end{document}